\documentclass[aps,preprint]{revtex4}%
\usepackage{amsfonts}
\usepackage{amsmath}
\usepackage{amssymb}
\usepackage{appendix}
\usepackage{graphicx}%
\begin{document}
\preprint{ }
\title[Gauge concepts]{Dual gauge concepts in electrodynamics, and new limitations on gauge invariance}
\author{H. R. Reiss}
\affiliation{Max Born Institute, Berlin, Germany}
\affiliation{American University, Washington, DC, USA}
\email{reiss@american.edu}

\begin{abstract}
Gauge invariance, a core principle in electrodynamics, has two separate
meanings. One concept treats the photon as the gauge particle for
electrodynamics. It is based on symmetries of the Lagrangian, and requires no
mention of electric or magnetic fields. The second concept depends directly on
the electric and magnetic fields, and how they can be represented by potential
functions that are not unique. A general proof that potentials are more
fundamental than fields serves to resolve discrepancies. Physical symmetries,
however, are altered by gauge transformations and strongly limit gauge
freedom. A new constraint on the form of allowable gauge transformations must
be introduced that applies to both gauge concepts.

\end{abstract}
\date[6 February 2022]{}
\maketitle

\section{Introduction}

From the time of Maxwell it was noted that electric $\mathbf{E}$ and magnetic
$\mathbf{B}$ fields could be represented as derivatives of scalar and 3-vector
potentials $\phi$ and $\mathbf{A}$ (that can be combined into a 4-vector
potential $A^{\mu}$). The association is not, in general, unique, and each
such representation is called a gauge. Another gauge concept arose from the
work of Yang and Mills \cite{yangmills} on the properties of isospin in strong
nuclear interactions, a precursor to a unified theory of gauge fields that
encompasses strong and weak nuclear interactions as well as electrodynamics.
We will be concerned here only with the electrodynamics component of gauge
fields, where the photon is the gauge particle carrying the interaction. Thus
there are two distinct gauge concepts in electrodynamics, often improperly
conflated: the first will be referred to as \textit{Maxwellian}, and the
second as \textit{Lagrangian,} since it depends on symmetries of the
Lagrangian density $\mathcal{L}_{EM}$ of the electromagnetic field.

Both disciplines employ the same concept of a gauge transformation of the
4-vector potential $A^{\mu}$, given by the rotation%
\begin{equation}
A^{\mu}\longrightarrow\widetilde{A}^{\mu}=A^{\mu}-\partial^{\mu}\Lambda,
\label{a}%
\end{equation}
where $\Lambda$ is any scalar function that satisfies the homogeneous wave
equation, and the Lorenz condition applies to both $A^{\mu}$ and
$\widetilde{A}^{\mu}$:%
\begin{equation}
\partial^{\mu}\partial_{\mu}\Lambda=0,\qquad\partial^{\mu}A_{\mu}%
=\partial^{\mu}\widetilde{A}_{\mu}=0. \label{1a}%
\end{equation}
These conditions are such that electric and magnetic fields are unchanged by
the transformation. \textit{Gauge invariance} refers to the requirement that a
change of gauge should not alter the outcome of calculations of electrodynamic interactions.

The conventional point of view is that preservation of electric and magnetic
fields is sufficient to identify gauge invariance. However, with the knowledge
that potentials are the determining factor in gauge transformations, the new
governing principle is that physical symmetries must be preserved in a gauge
transformation. This is much more demanding than the traditional concept. An
important alteration in the original notion of gauge invariance is the
symmetry required to sustain the propagation property of plane waves. It is
shown that dependence on electric and magnetic fields is not sufficient to
sustain propagation, and the recasting to a potentials requirement strongly
constrains the possible class of equivalent gauges. The presence of a Coulomb
binding potential in addition to plane waves leaves no freedom at all in the
selection of gauge.

Gauge invariance is regarded as a bedrock principle in electrodynamics.
Maxwellian gauge concepts can be traced back to Maxwell's equations of 1865
\cite{maxwell}, and the Lagrangian form to the Yang-Mills paper of 1954
\cite{yangmills}. Despite these venerable origins, the fact that two
inequivalent forms of gauge invariance exist seems never to have been studied critically.

The following Section describes briefly the Lagrangian formulation, which has
remained free of controversy, unlike the Maxwellian formalism treated in much
more detail in Section III. However, a new limitation on gauge transformations
must be added to both gauge concepts to preserve the propagation property of
plane waves.

Difficulties with the meaning and application of the Maxwellian formalism
continue to the present day.

A proof, of much broader scope than that of Aharonov-Bohm, demonstrates that
potentials provide more information than electric and magnetic fields, and
that the fields can give erroneous predictions. When the Maxwellian
field-based discipline is altered to be in terms of potentials, then there is
no conflict with Lagrangian gauge transformations for plane wave fields. The
Lagrangian formalism examines only the photon 0field, equivalent to plane-wave
fields, whereas the Maxwellian formalism is not limited in this way, and thus
has a much broader scope.

Although gauge equivalence remains valid, it is shown that the concept of
\textit{gauge freedom} is significantly restricted due to changes in physical
symmetries that follow from gauge transformations.

The status of potentials as the primary quantities in electrodynamics
provides\ insights as to why the tunneling model for atomic ionization is
useful only for some ranges of parameters, and why an analogous situation
exists for the concept of the local constant field approximation (LCFA).

Electromagnetic quantities are expressed in Gaussian units.

\section{Lagrangian Formalism}

The contrast between the Maxwellian and Lagrangian approaches becomes evident
when the electromagnetic Lagrangian density is displayed:%
\begin{equation}
\mathcal{L}_{EM}=-F^{\mu\nu}F_{\mu\nu},\qquad F^{\mu\nu}=\partial^{\mu}A^{\nu
}-\partial^{\nu}A^{\mu}.\label{b1}%
\end{equation}
$\mathcal{L}_{EM}$ is always exactly preserved in the gauge transformation
(\ref{a}), which explains why the Lagrangian formalism is free of controversy.

Another essential point is that the Lagrangian formalism depends entirely on
the 4-vector potential $A^{\mu}$, so it can be explored without ever
mentioning $\mathbf{E}$ or $\mathbf{B}$ fields. This is in contrast to the
Maxwellian formalism, where the core consideration is how $\mathbf{E}$ and
$\mathbf{B}$ fields are related to potentials.

A unfortunate consequence of ignoring this disparity in gauge formulations is
exemplified by a recent pair of papers where an examination of Maxwellian
properties \cite{davis1} is deemed invalid on the grounds of the great success
of the Lagrangian formalism \cite{hees}.

A previously unexplored limitation must be added to the standard requirements
for valid gauge transformations to preserve the propagation property of plane
waves. This is explained in Section IV.

\section{Maxwellian Formalism}

Problems with the Maxwellian formalism can be divided into two categories. One
is simply an improper application of the formalism. The second category of
problems refers to fundamental failures to satisfy basic principles of physics
when fields are regarded as more fundamental than potentials.

We start with a quick review of fundamentals.

\subsection{Basic Premises}

From early stages of the study of electromagnetism in vacuum, the electric
field $\mathbf{E}$ and the magnetic field $\mathbf{B}$ were viewed as basic
measures. The connection of $\mathbf{E}$ and $\mathbf{B}$ to scalar and vector
potentials $\phi$ and $\mathbf{A}$ is not unique, so preservation of
$\mathbf{E}$ and $\mathbf{B}$ in gauge transformations was regarded as
fundamental. This was not called into question until the Aharonov-Bohm process
\cite{ehrsid,ab,berry} was noted (to be discussed below), but some authorities
\cite{jackokun} still maintain that the fields $\mathbf{E}$ and $\mathbf{B}$
remain fundamental.

\subsection{Improper Inferences}

The often-controversial aspects of the Maxwellian formalism to be discussed
below do not occur at all in the Lagrangian formalism.

\subsubsection{Length gauge}

The improprieties to be reviewed are to be found principally in the Atomic,
Molecular, and Optical (AMO) physics community. The Appendix recounts a brief
history of the predilection of the AMO community to favor the length gauge,
with some saying that it is the only \textquotedblleft
correct\textquotedblright\ gauge.

\subsubsection{Dipole approximation vs. long-wavelength approximation}

The dipole approximation (DA) is taken in the AMO community to mean a complete
neglect of the magnetic component of a plane wave (PW). It is sometimes
confused with the long-wavelength approximation (LWA), which simply implies
that the wavelength of a PW is much larger than the extent of the atom. The
LWA does not mean that the magnetic field is neglected. The distinction is important.

The DA is a much stronger constraint than the LWA because the DA sets
$\mathbf{B}=0$ from the outset. If $\mathbf{A}\left(  t,\mathbf{r}\right)  $
is used in its full spacetime form to find $\mathbf{B}\left(  t,\mathbf{r}%
\right)  =\mathbf{\nabla\times A}\left(  t,\mathbf{r}\right)  $, and then the
LWA is applied to $\mathbf{B}$, a $\mathbf{B}\left(  t\right)  $ is found that
becomes increasingly important as intensity increases or as frequency
decreases, with $\mathbf{B}$ becoming significant with existing laboratory
parameters \cite{hr101,hrtun}. That is, the two actions of evaluating
$\mathbf{\nabla\times A}$ and applying the LWA are not commutative.

If the DA is employed in its usual form with both $\mathbf{B}=0$ and the LWA,
then $\omega\rightarrow0$ gives only a static electric field. On the other
hand, if the LWA is applied after $\mathbf{B}\left(  t,\mathbf{r}\right)
=\mathbf{\nabla\times A}\left(  t,\mathbf{r}\right)  $ is evaluated, the
result is a low-frequency PW that propagates at the speed of light. The
outcome of the two procedures: a static electric field vs. a field that
propagates at the speed of light, could hardly be more different.

\subsubsection{Oscillatory electric vs. plane wave}

An important error arises when the condition $\lambda\gg1$ a.u. is viewed as
the only constraint on the DA, where $\lambda$ is the wavelength of the PW
field. This approach makes it natural to suppose that very long wavelengths
$\lambda$ (i.e. low frequencies $\omega$) are far removed from any possibility
of DA failures, and low frequency situations are then assumed to be governed
by what is called \textquotedblleft adiabaticity\textquotedblright,
approaching constant electric-field behavior as $\omega\rightarrow0$. An
unfortunate effect was that analytical approximations for strong-field
processes were judged deficient unless they approached constant-field results
in the low frequency limit. See, for example, Ref. \cite{joachain}.

The actuality is that oscillatory electric-field effects lose all resemblance
to plane-wave field effects as $\omega\rightarrow0$ \cite{hr101,hrtun},
contradicting the adiabaticity argument. A plane-wave field propagates at the
speed of light for any frequency, no matter how low, with $\mathbf{E}$ and
$\mathbf{B}$ fields of equal magnitude. By contrast, the DA version infers an
oscillatory electric field that approaches a static limit with no magnetic
field at all.

\subsection{Equations of motion}

A paradox with the Maxwellian formalism was examined in the 1960s. All
quantum-mechanical equations of motion for particles in interaction with
electromagnetic fields have that interaction represented by potentials. The
prevailing view was that electric and magnetic fields are fundamental, but all
attempts \cite{mandelstam,dewitt,belinfante,levy,rohrlich,priou} to express
the Schr\"{o}dinger equation directly in terms of electric and magnetic fields
led to nonlocal theories. In retrospect this was an inevitable result.
Although fields are found from potentials by differentiation, which is a local
procedure, potentials are found from fields by integration, which is nonlocal.
Thus, any attempt to replace the potentials in quantum equations of motion by
fields will inevitably introduce nonlocality. It applies as well as for all
classical equations of motion except for the Newtonian form.

\section{Potentials are fundamental; fields are secondary}

The fact that equations of motion are expressed in terms of potentials, which
therefore govern the properties of the solutions of those equations, is a
strong indicator of the primacy of potentials. There are, however, even more
direct indicators: the Aharonov-Bohm effect and the properties of PWs
propagating in the vacuum.

\subsection{Aharonov-Bohm effect}

A major development in the matter of \ potentials vs. fields is the 1959 paper
of Aharonov and Bohm (AB) \cite{ehrsid,ab,berry}. AB considered the
quantum-mechanical experiment in which a beam of electrons is deflected as it
passes over a closed solenoid. There is a magnetic field within the solenoid,
but no field outside. There is, however, a potential outside, so it must be
the potential that gives rise to the deflection, meaning that the potential is
more fundamental than the field. Experimental verification of the deflection
was achieved in 1986 \cite{tonomura}.

Many investigators accept that the AB effect is a universal proof that
potentials are more fundamental than fields, but others regard it as a
quantum-only effect, or that the effect is too particularized to be taken as a
general proof of the primacy of potentials over fields. Jackson and Okun
\cite{jackokun} maintain that it is possible to salvage the long-standing
conviction that fields are physical. Part of their argument is that
$\mathbf{E}$ and $\mathbf{B}$ can be measured in the laboratory, but $A^{\mu}$
cannot. However, the ponderomotive potential $U_{p}$ is proportional to
$A^{\mu}A_{\mu}$, and $U_{p}$ is the fundamental measure of the
nonperturbative interaction of a charged particle with a plane-wave field
\cite{hr62,hrmass,hr100}. The ponderomotive potential is observable as the
cause of channel closings \cite{hr80,muller,hrepjd}.

\subsection{Plane waves}

The study of the properties of PWs (also known as transverse fields,
propagating fields, sourceless fields, laser fields, or simply
\textquotedblleft light\textquotedblright) is fundamental in classical
electrodynamics, quantum electrodynamics, and astrophysics. To satisfy the
Einstein basic principle of relativity that the speed of light is the same in
all inertial frames of reference \cite{einstein}, the $A^{\mu}$ that describes
the PW can depend on the spacetime 4-vector $x^{\mu}$ only as the scalar
product with the propagation 4-vector $k^{u}$:%
\begin{equation}
A^{\mu}\left(  \varphi\right)  ;\qquad\varphi\equiv k^{\nu}x_{\nu}=\omega
t-\mathbf{k\cdot r} \label{d3}%
\end{equation}

It is possible to introduce a gauge transformation from a set of potentials
meeting the Einstein condition to\ a new set of potentials that violate it.
Starting with a 4-vector potential of the form given in Eq. (\ref{d3}), a
gauge transformation can be generated \cite{hr79,hr100} by the function%
\begin{equation}
\Lambda=A^{\mu}x_{\mu}. \label{e}%
\end{equation}
Equation (\ref{a}) takes the form%
\begin{equation}
\widetilde{A}^{\mu}=-k^{\mu}x^{\nu}(d_{\varphi}A_{\nu})\text{ \ \ or
\ \ }\widetilde{A}^{\mu}=-\frac{k^{\mu}}{\omega/c.}\mathbf{r\cdot E}\left(
\varphi\right)  , \label{f}%
\end{equation}
where the second expression in Eq. (\ref{f}) is the result if the initial
gauge before transformation is the Coulomb gauge. Neither of the forms in Eq.
(\ref{f}) can be descriptive of a propagating field. The occurrence of
$x^{\nu}$ in isolation from $k_{\nu}$ in the first expression and the
appearance of $\mathbf{r}$ in the second one, requires that an origin of
coordinates must be specified. This is a clear violation of the Einstein
principle. Nevertheless, the gauge transformation generated by Eq. (\ref{e})
satisfies the Lorenz condition and the homogeneous wave equation. It is a
valid gauge transformation, meaning that the $\mathbf{E}$ and $\mathbf{B}$
fields in the transformed gauge are identical to those in the initial gauge.

The 4-potential $A^{\mu}\left(  \varphi\right)  $ is an appropriate form for
the description of a propagating electromagnetic field, but the potentials in
Eq. (\ref{f}) cannot describe a propagating field. A knowledge of the
potentials is necessary to distinguish the physical from the unphysical
situation; the $\mathbf{E}$ and $\mathbf{B}$ fields are unable to make the distinction.

\subsection{Consequences of potentials over fields}

Among the consequences that flow from the primacy of potentials, two will be
mentioned here because they are important for currently active research.

\subsubsection{Effects of plane waves vs. effects of oscillatory electric
fields}

When the effects of plane waves are treated with the DA, this is tantamount to
replacing a propagating field by one that is simply an oscillatory electric
field. That is, a wave that propagates with the speed of light is replaced by
a simple oscillatory electric field that has zero speed of propagation.

Both the DA and the LWA, as used in AMO physics, are subject to an upper limit
on frequency equivalent to the wavelength limit $\lambda\gg1a.u.$ As the
frequency declines, effects predicted by the DA are close to those of the LWA,
but that correspondence vanishes when the coupling of the electric and
magnetic components of a PW field induces an oscillation of amplitude
$\beta_{0}$ \cite{ss} in the propagation direction approaching $1a.u.$, where
\begin{equation}
\beta_{0}=\frac{U_{p}}{2mc\omega}\frac{1}{\left(  1+2U_{p}/mc^{2}\right)
}\approx\frac{U_{p}}{2mc\omega}.\label{g}%
\end{equation}
The quantity $U_{p}$ is the ponderomotive energy of a charged particle in the
field. As the lower limit on frequency approaches the value in Eq. (\ref{g}),
the difference between DA and LWA predictions diverge. Since $U_{p}$ has the
0frequency dependence $1/\omega^{2}$, then%
\begin{equation}
\beta_{0}\backsim1/\omega^{3}.\label{g2}%
\end{equation}
As $\omega\rightarrow0$, the DA approaches static electric field behavior,
whereas the LWA tends towards relativistic behavior. The onset of this extreme
divergence is very rapid, as Eq. (\ref{g2}) demonstrates.

The in-phase coupling between electric and magnetic fields in a PW produces a
force (radiation pressure) in the direction of propagation of the PW,
producing the classical figure-8 motion \cite{ss} in linearly polarized light.
In qualitative terms, calculations within the DA are in good agreement with
experiments until the point where effects of coupling between electric and
magnetic components become discernable \cite{hr101,hrtun}. Several
laboratories now have the capability to explore non-DA behavior.

\subsubsection{Effects of plane waves vs. effects of constant crossed fields.}

Constant crossed fields are an example showing the inability of electric and
magnetic fields to uniquely define an electromagnetic environment. One
approach to the significance of constant crossed field comes from an
examination of the Lorentz invariants of electromagnetic fields. Two Lorentz
invariants can be formed from the electromagnetic field tensor of Eq.
(\ref{c}):%
\begin{equation}
F^{\mu\nu}F_{\mu\nu}\thicksim\mathbf{E}^{2}-\mathbf{B}^{2},\qquad F^{\mu\nu
}\mathcal{F}_{\mu\nu}\thicksim\mathbf{E\cdot B,} \label{h}%
\end{equation}
where $\mathcal{F}_{\mu\nu}$ is the dual to the field tensor. When both of
these invariants have the value zero the field is called a \textit{null
field}. The most important example is a PW field, where
\begin{equation}
\left\vert \mathbf{E}\right\vert =\left\vert \mathbf{B}\right\vert
,\qquad\mathbf{E\perp B}. \label{i}%
\end{equation}
The propagating field requires only the condition of Eq. (\ref{d3}), but it is
also possible to produce constant $\mathbf{E}$ and $\mathbf{B}$ fields that
satisfy the null field conditions of Eq. (\ref{i}). Many authors have
concluded that the constant-crossed-field configuration provides a simple way
to assess the effects of a propagating field. This was initially proposed in
Refs. \cite{nikrit,ritus}, and has been explored by many authors as the LCFA.

Another line of reasoning is to regard a low-frequency laser field as having
so long a wavelength as compared to an object of atomic size that the
$\mathbf{E}$ and $\mathbf{B}$ fields approximate static conditions in the
interaction region. A good way to appraise this conjecture is to compare the
effects of a constant crossed field with those of a propagating field of
extremely long wavelength. This can be done in the context of a practical
example employed in Ref. \cite{hrnotun}: an extremely-low-frequency (ELF)
radio transmitter employed by the US Navy \cite{sanguine} to communicate with
deeply submerged submarines at distances of thousands of kilometers from the
transmitting antenna. The system operated at 76 Hz, corresponding to a
wavelength of about 4,000 km. This is about 1/3 the diameter of the Earth. On
a laboratory scale, such a radio wave would appear to have $\mathbf{E}$ and
$\mathbf{B}$ values constant to very high accuracy. As has been shown above,
it is not possible to describe static $\mathbf{E}$ and $\mathbf{B}$ components
in terms of the propagating field potential of the form of Eq. (\ref{d3}).
Appropriate potentials for static $\mathbf{E}_{0}$ and $\mathbf{B}_{0}$ fields
are%
\begin{equation}
\phi=-\mathbf{r\cdot E}_{0},\qquad\mathbf{A}=-\frac{1}{2}\mathbf{r\times
B}_{0}. \label{j}%
\end{equation}
There is no gauge transformation that can relate the potentials obeying Eq.
(\ref{j}) to those satisfying Eq. (\ref{d3}). The primacy of potentials over
fields means that the potentials of Eq. \ref{j} represent an entirely
different physical environment from potentials in the form of Eq. (\ref{d3}).

When a solution of the Schr\"{o}dinger equation is sought, the potentials
determine the outcome, so the potentials of Eq. (\ref{j}) do not exhibit the
properties of PWs. This failure is not evident in terms of electric and
magnetic \ fields. It is only the potentials that determine a correct result.

As in the case of the near equivalence of the DA to true plane-wave effects in
a limited range of parameters, it has been found that at high intensities the
LCFA mimics PW behavior satisfactorily \cite{dipiazza1,dipiazza2}.

\section{Constraints on gauge freedom}

\subsection{Plane waves}

The usual restrictions in a gauge transformation are that the generating
function should be a scalar function that satisfies the homogeneous wave
equation, and that the initial and transformed 4-vectors $A^{\mu}$ and
$\widetilde{A}^{\mu}$ should obey the Lorenz condition, as shown in Eq.
(\ref{1a}). These are not sufficient to preserve all necessary symmetries.

For plane waves, the property stated in Eq. (\ref{d3}) that both $A^{\mu}$ and
$\widetilde{A}^{\mu}$ are restricted to have spacetime dependence only in
combination with the propagation 4-vector $k^{\mu}$ as the product
$\varphi=k^{\mu}x_{\mu}$, means that $\partial^{\mu}\Lambda$ must also have
this property. This gives%
\begin{equation}
\partial^{\mu}\Lambda\left(  \varphi\right)  =\partial^{\mu}\left(
\varphi\right)  \frac{d}{d\varphi}\Lambda\left(  \varphi\right)  =k^{\mu
}\Lambda^{\prime}\left(  \varphi\right)  . \label{n}%
\end{equation}
That is, any change in the 4-vector potential resulting from a gauge
transformation has to lie on the lightcone.

This is a very strong constraint that applies both in\ classical and quantum
electrodynamics. It applies to plane waves, which are at the heart of
electrodynamics, yet it is not mentioned in standard texts on electrodynamics.

\subsection{Plane waves + Coulomb potential}

A common laboratory situation occurs when an electron subjected to a Coulomb
binding force is irradiated with a laser beam. The strong constraint (\ref{n})
imposed for the description of the laser beam becomes even stronger when the
binding potential must be included. The Dirac and Klein-Gordon equations must
now incorporate both interactions. This is done by adding the scalar potential
$V\backsim1/r$ to the time component of the plane wave potential:%
\begin{equation}
A^{0}=A_{PW}^{0}+V. \label{o}%
\end{equation}
The Klein-Gordon equation, with time and space parts separated is%
\begin{equation}
\left[  \left(  \frac{i\hslash}{c}\partial_{t}-\frac{q}{c}A^{0}\right)
^{2}-\left(  ii\hslash\mathbf{\nabla}-\frac{q}{c}\mathbf{A}\right)
^{2}-\left(  mc\right)  ^{2}\right]  \psi=0. \label{p}%
\end{equation}
When Eq. (\ref{o}) is inserted into the first parenthesis in Eq. (\ref{p}),
and the indicated square is performed, there will be a cross term $A_{PW}%
^{0}V$ that causes a problem in the nonrelativistic reduction of the
Klein-Gordon equation to the Schr\"{o}dinger equation \cite{schoene,schiff}
due to the singularity of the Coulomb potential.

\subsection{Beyond plane waves}

The Lagrangian formalism is in terms of transverse fields, but the scope of
the Maxwellian is much broader. Transforming gauge means that there will be a
change in the functional dependence of the potentials. A change in functional
form of the potentials means that conservation principles will be altered. In
the case of plane waves it was shown that the condition in Eq. (\ref{n}) must
be satisfied in order to maintain the propagation property of plane waves. The
general effect of changing the form of the potentials can be seen even in the
elementary problem of an electron in a constant electric field \cite{hr100}.
That case, from Ref. \cite{hr100} is replicated in the Appendix to illustrate
the problem in simple terms.

The overall conclusion is that gauge transformations do not automatically
preserve the outcome of a physical calculation. In general, additional
constraints must be satisfied, and these extra requirements may be very restrictive.

\section{Overview}

The \textquotedblleft traditional\textquotedblright\ concept of gauge
invariance was based on the belief that electric and magnetic fields are the
physical determinants of an electrodynamic environment. Gauge invariance was
based on preservation of the fields. A basic flaw in this argument is that the
Schr\"{o}dinger equation and other equations of motion are based on
potentials, so that a change of gauge will alter the solution of an
electrodynamic problem. An example is shown here where a valid gauge
transformation involving plane waves is such that the propagation property of
a set of potentials is lost as a result of the gauge transformation. The
solution of the transformed equation of motion is unphysical even though the
gauge transformation preserves the electric and magnetic fields. The
resolution of this apparent dilemma is that the potentials determine the
solution, not the fields.

With this alteration, there is then no conflict with gauge invariance based on
electromagnetism as the gauge field within the Standard Model, where only the
potentials are involved and electric and magnetic fields need not be mentioned.

The example used to arrive at this conclusion is Eq. (\ref{e}), and this gauge
transformation introduces a violation of the condition that plane waves
propagate with the speed of light in any Lorentz frame. An additional
condition must be added to the customary conditions governing gauge
transformations. This new constraint requires that all gauge transformations
involving plane waves must lie on the lightcone. That is, they must have the
same spacetime dependence as the propagation 4-vector $k^{\mu}$.

\section{Appendix: Special status of the length gauge}

Controversy about the Maxwellian formalism began with a 1952 paper by Lamb
\cite{lamb} and continues to the present \cite{davis1,hees,heras,davis2}, 70
years later.

In Atomic, Molecular, and Optical (AMO) physics, two gauges are conventionally
employed: the \textquotedblleft length gauge\textquotedblright, with
interaction Hamiltonian
\begin{equation}
H_{I}=-q\mathbf{r\cdot E}\left(  t\right)  , \label{c}%
\end{equation}
and the \textquotedblleft velocity gauge\textquotedblright\ with
\begin{equation}
H_{I}=-\frac{q}{mc}\mathbf{A}\left(  t\right)  \mathbf{\cdot p+}\frac{q^{2}%
}{2mc^{2}}\mathbf{A}^{2}\left(  t\right)  . \label{d1}%
\end{equation}
Both expressions employ the dipole approximation (DA), routinely adopted in
the AMO community, and defined to mean omitting spatial dependence in
$\mathbf{E}$ and $\mathbf{A}$. This leads to $\mathbf{B}=0,$ thus losing any
possibility of treating the propagation property of plane-wave fields that is
vital in the exploration of laser-induced phenomena. Plane-wave fields in the
vacuum have equal magnitudes of electric and magnetic fields:%
\begin{equation}
\left\vert \mathbf{E}^{PW}\right\vert =\left\vert \mathbf{B}^{PW}\right\vert ,
\label{d2}%
\end{equation}
so setting $\mathbf{B}^{PW}=0$ is questionable in view of the Maxwellian
demand that $\mathbf{E}$ and $\mathbf{B}$ are both fundamental. Coupling of
the electric and magnetic components of a PW becomes vitally important when
field intensity is high and/or when field frequency is low \cite{hr101,hrtun}.

Equation (\ref{c}) can be descriptive of an oscillatory electric field, but it
is a scalar potential, which is fundamentally inadequate for the
representation of a vector field such as that of a laser \cite{hrepjd55}.

In 1952, Lamb \cite{lamb} did a second-order perturbative line-shape
calculation and found a plausible result in the length gauge, but an
unacceptable asymmetry in the velocity gauge. Lamb viewed this failure of
gauge invariance as perplexing, but Power and Zienau \cite{power} asserted
that if length and velocity gauges give different answers, the length gauge is
correct.~They leave unresolved the dilemma about how gauge-equivalent
expressions can lead to different physical predictions. Fried demonstrated
\cite{fried} in 1973 that Lamb had erred in his velocity-gauge line shape
calculation and, when properly treated, the gauge conflict is removed; but
this resolution of the difficulty was not acknowledged in the AMO community.
K. H. Yang \cite{yang} went further than Power and Zienau by claiming to prove
that the length gauge is a fundamental gauge with which all calculations must
begin, followed by a gauge transformation if any other gauge is to be
employed. This viewpoint gained much support
\cite{kobesmirl,kobeyang,schlicherbeck,lambschlicher,heidelberg,bergues1,bergues2}
even for problems with strong laser fields, despite the fact that all
plane-wave fields like those of a laser beam are vector fields. Vector fields
in interaction with matter cannot properly be represented by a scalar potential.

A questionable belief widespread among defenders of the primacy of the
interaction Hamiltonian of Eq. (\ref{c}) is the assumption that a gauge
transformation is a unitary transformation. If this was true then%
\begin{equation}
U\mathbf{r\cdot E}U^{-1}=\mathbf{r\cdot E}, \label{e1}%
\end{equation}
since there are no differential operators in $\mathbf{r\cdot E}$. This leads
to the concept of a \textquotedblleft gauge-invariant gauge\textquotedblright%
\ or a \textquotedblleft gauge-invariant formulation\textquotedblright%
\ \cite{heidelberg}. Apart from the transparent illogic (e.g. if one can
transform into a gauge-invariant gauge, then the inverse gauge transformation
could not exist), there is the fact that a gauge transformation is not, in
general, unitary\ \cite{hr100}. Starting from the well-known form-invariance
of the Schr\"{o}dinger equation, it follows that the gauge-transformed
Hamiltonian $\widetilde{H}$ is \cite{hr100}%
\begin{equation}
\widetilde{H}=UHU^{-1}-i\hslash\left(  \partial_{t}U^{-1}\right)  , \label{f1}%
\end{equation}
meaning that any time-dependence in the generating function for the gauge
transformation precludes unitarity. This holds true even if $\mathbf{E}$ is time-independent.

For example, consider the elementary problem where a charged particle is
subjected to a constant electric field $\mathbf{E}_{0}$ \cite{hr100}. The
field can be specified by the potentials.%
\begin{equation}
\phi=-\mathbf{r\cdot E}_{0},\qquad\mathbf{A}=0. \label{g1}%
\end{equation}
The problem is time-independent, meaning that energy is conserved. A gauge
transformation to the gauge where%

\begin{equation}
\widetilde{\phi}=0,\qquad\widetilde{\mathbf{A}}=-ct\mathbf{E}_{0} \label{h1}%
\end{equation}
is accomplished with the generating function%
\begin{equation}
\Lambda=-ct\mathbf{r\cdot E}_{0}. \label{i1}%
\end{equation}
The potentials are now independent of spatial coordinates, so that momentum is
conserved. Two things are manifest in this example: conservation principles
are altered, and time dependence of $U$ occurs even in an initially
time-independent problem.

This constant electric field problem is so elementary that if a particle
trajectory is sought, the same classical trajectory is found from either Eq.
(\ref{g1}) or Eq. (\ref{h1}), albeit with different procedures. That is
because the problem is such that both energy and momentum are conserved. In
other contexts, like the plane wave case, new constraints must be observed in
a change of gauge.

\end{document}